\documentclass[a4paper,11pt]{article}
\usepackage{pos}
\usepackage{subcaption}
\usepackage{tabularray}
\usepackage[bottom]{footmisc}
\usepackage{gensymb,multirow,wrapfig,lineno,enumitem}

\title{New full-sky studies of the distribution of ultra-high-energy cosmic-ray arrival directions}
\ShortTitle{New full-sky studies of the distribution of UHECR arrival directions}

\author*[a]{A.~Gálvez Ureña}
\author{L.~Anchordoqui}
\author{M.~Bianciotto}
\author{P.~Biermann}
\author{T.~Bister}
\author{J.~Biteau}
\author{L.~Caccianiga}
\author{O.~Deligny}
\author{L.~Deval}
\author{A.~di~Matteo}
\author{U.~G.~Giaccari}
\author{G.~Golup}
\author{R.~Higuchi}
\author{J.~Kim}
\author{M.~Kuznetsov}
\author{J.~P.~Lundquist}
\author{F.~M.~Mariani}
\author{G.~Rubtsov}
\author{P.~Tinyakov}
\author{F.~Urban}
\onbehalf{for the Pierre Auger Collaboration$^{b,1}$ and the Telescope Array Collaboration$^{c,2}$}

\affiliation[a]{CEICO—FZU, Institute of Physics of the Czech Academy of Sciences, Na Slovance 1999/2, 182 00 Prague, Czech Republic}
\emailAdd{urena@fzu.cz}

\affiliation[b]{Observatorio Pierre Auger,
Av.\ San Martín Norte 304, 5613 Malargüe, Argentina}
\note{Full author list: \url{https://www.auger.org/archive/authors_icrc_2025.html}}
\emailAdd{spokespersons@auger.org}

\affiliation[c]{Telescope Array Project, 201 James Fletcher Bldg., 115 S. 1400 E., Salt Lake City, UT 84112-0830,
USA}
\note{Full author list: \url{http://telescopearray.org/index.php/research/collaborators}}
\emailAdd{ta-icrc@cosmic.utah.edu}

\makeatletter
\newcommand*{\ssymbol}[1]{\ensuremath{\@fnsymbol{#1}}}
\makeatother

\abstract{Ground-based full-sky studies of the angular distribution of arrival directions of ultra-high-energy cosmic rays require combining data from different observatories, such as the Pierre Auger Observatory (Auger) and the Telescope Array (TA), because no single array can cover all declinations. A working group comprising members from the Auger and TA collaborations has been tasked with performing such studies for more than a decade and has found several indications of full-sky anisotropies. Here, we update the results for the large- and medium-scale anisotropy analyses using the latest data from TA, which include corrections for daily and yearly atmospheric effects in data for large-scale anisotropies and looser selection criteria in data for medium-scale anisotropies. We extend the latter one by considering two more galaxy catalogues, consisting of jetted or all AGNs. Finally we also introduce a new angular harmonic space analysis that allows us to measure both the auto-correlation and cross-correlation with all catalogues for all multipoles independently ($\ell_\text{max} = 20$ in this work) and scanning the energy threshold.}

\FullConference{39th International Cosmic Ray Conference (ICRC 2025)\\
 15–24 July 2025 \\
Geneva, Switzerland\\}

\begin{document}
\maketitle
\section{Introduction}
The most energetic particles ever measured are ultra-high-energy cosmic rays (UHECRs), which are nuclei coming from space with energies above $1\,\mathrm{EeV} = 10^{18}\,\mathrm{eV}$. At such high energies, UHECRs interact with the cosmic microwave background (CMB) and the extragalactic background light (EBL) and quickly lose energy. This is why the sources of UHECRs must be nearby. Despite this, due to intervening magnetic fields, we do not expect the arrival direction of the particles to point back to the origin. As a result, direct association of an UHECR to its source is very challenging.
 
We look at large-scale anisotropies such as the dipole or quadrupole, where the effect of the magnetic fields is expected to be smaller \cite{diMatteo2018}. To study larger scales, having full-sky coverage is extremely beneficial, since partial sky coverage could introduce strong degeneracies between the first few multipoles. We achieve this by combining data from the Pierre Auger Observatory (Auger) in Argentina and the Telescope Array (TA) in Utah, USA\@. Together they cover all declinations. Declinations~$-15.7\degree < \delta < +44.8\degree$ are covered by both observatories, allowing us to cross-calibrate the data. 

In this proceeding, we update the results of the large-scale and medium-scale anisotropy analyses that have been presented before \cite[and refs.\ therein]{Auger+TA_UHECR2024}. Additionally, we extend the medium-scale anisotropy analysis to two more galaxy catalogs previously used in Auger-only analyses \cite{Auger_MSA}, consisting of AGNs observed from the \textit{Fermi}-LAT and \textit{Swift}-BAT telescopes. Finally, we introduce a new angular harmonic-space analysis to search for anisotropies at large and medium scales at the highest energies.

\section{The datasets}
\begin{wrapfigure}{R}{0.5\textwidth}
    \centering
    \includegraphics[page=4]{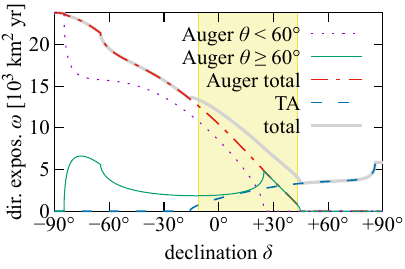}
    \caption{Directional exposures of the datasets we used for large-scale anisotropies (those for medium-scale anisotropies have the same shapes and slightly higher normalizations)}
    \label{fig:exposure}
\end{wrapfigure}
We use the latest datasets available, including events detected in Auger from 1~January 2004 to 31~December 2022 for Auger and in TA from 11~May 2008 to
10~May 2024. The Auger datasets are the same as in our previous work \cite{Auger+TA_UHECR2024}: one with stricter cuts \cite{Auger_LSA} for large-scale anisotropies, with an exposure of~$123\,000\,\mathrm{km^2\,sr\,yr}$, and one with looser cuts \cite[but with a longer time period]{Auger_MSA} for medium-scale anisotropies at the highest energies, whose exposure is~$135\,000\,\mathrm{km^2\,sr\,yr}$. As for TA, for large-scale anisotropies we use the same dataset as in our previous study \cite{Auger+TA_UHECR2024}, with strict cuts \cite[but with a longer time period]{TA_LSA} and an effective exposure (accounting for energy resolution effects) of $19\,500\,\mathrm{km^2\,sr\,yr}$, except that here for the first time we correct energy measurements for yearly and daily variations in the air density.\footnote{
    Namely, $E_\text{corr} = E_\text{raw}\times \left(1 - 3.5\bigl(\frac{\rho}{1.042\,\mathrm{kg/m^3}} - 1\bigr)\right)^{-1/1.7}$, where $\rho$ is the air density at the TA site obtained from the Global Data Assimilation System every three hours.
} Additionally, for medium-scale anisotropies and the angular harmonic analyses, we use looser cuts \cite[but with a longer time period]{TA_MSA} for events with $E_\text{TA} > 57\,\text{EeV}$, with 52 extra events and $5\,800\,\mathrm{km^2\,sr\,yr}$ extra exposure compared to the strict-cut dataset. 
See \autoref{fig:exposure} for the exposure of the datasets as a function of declination.

\begin{wrapfigure}[18]{r}{0.5\textwidth}
    \centering
    \includegraphics[]{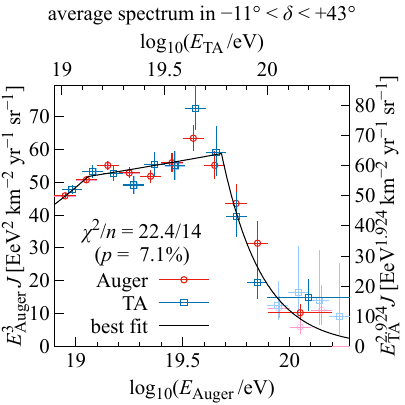}
    \caption{The spectrum fit to the two datasets in order to obtain a mapping between Auger and TA energy scales.  The highest-energy bins of each dataset (denoted by thin pale lines) are combined into one larger bin (denoted by regular lines) to ensure the probability distribution can be approximated as log-normal.}
    \label{fig:spectrum}
\end{wrapfigure}
UHECR energy measurements are affected by sizeable systematic uncertainties ($\pm$14\% for Auger and  $\pm$21\% for TA\@). This must be carefully considered, as it may have a significant impact on the results, particularly on the dipole component. We use the events in the common declination band to cross-calibrate both datasets, as in~\cite{Auger+TA_ICRC2021}.\footnote{This cross-calibration is optimized for anisotropy studies and should not be used outside of the scope of this analysis.} This is done by fitting 
the spectrum model seen in \autoref{fig:spectrum} and the parameters $\alpha$ and $\beta$ in:
\begin{align*}
    E_\text{Auger}/10\,\mathrm{EeV} &= \mathrm{e}^\alpha          (E_\text{TA}/10\,\mathrm{EeV})^\beta, \\
    E_\text{TA}/10\,\mathrm{EeV} &= \mathrm{e}^{-\alpha/\beta} (E_\text{Auger}/10\,\mathrm{EeV})^{1/\beta}.
\end{align*}
The values obtained are~$\alpha = -0.150 \pm 0.011$ and $\beta = 0.962 \pm 0.016$, resulting in \begin{align*}
       E_\text{TA} = 10\,\mathrm{EeV} &\leftrightarrow E_\text{Auger} = 8.60\,\mathrm{EeV}, \tag{$E_1$}\\
    E_\text{Auger} = 16\,\mathrm{EeV} &\leftrightarrow E_\text{TA} = 19.1\,\mathrm{EeV}, \tag{$E_2$}\\
    E_\text{Auger} = 32\,\mathrm{EeV} &\leftrightarrow E_\text{TA} = 39.2\,\mathrm{EeV}. \tag{$E_3$}
\end{align*}

\begin{figure}[b]
    \centering
    \includegraphics[width=0.33\linewidth,page=1]{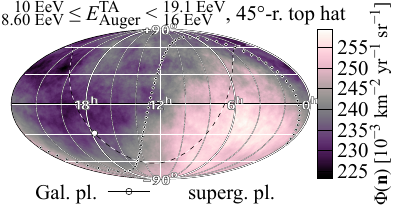}%
    \includegraphics[width=0.33\linewidth,page=2]{Figures/top_hat-45.pdf}%
    \includegraphics[width=0.33\linewidth,page=3]{Figures/top_hat-45.pdf}
    \caption{The flux distribution of the dataset we use for large-scale anisotropies (equatorial coordinates, Mollweide projection), smoothed by a $45\degree$-radius top-hat window}
    \label{fig:flux}
\end{figure}
In the dataset used for large-scale anisotropies, there are $43\,600$~Auger events with~$E \ge E_1$, of which $13\,027$ with~$E \ge E_2$ and 2\,739 with~$E \ge E_3$, and $6\,611$~TA events with~$E \geq E_1$, of which 1\,967 with~$E \ge E_2$ and 484 with~$E \ge E_3$.  The flux distribution in this dataset in the three energy bins is shown in \autoref{fig:flux}.

As previously mentioned, for the intermediate scale anisotropies and the harmonic analyses we use the TA loose-cut data for $E_\text{TA} \geq 57 \,\text{EeV}$. However, since this dataset does not have available atmospheric corrections, we use the energy calibration from \cite{Auger+TA_UHECR2024}, with $\alpha=-0.159$, $\beta=0.954$, and $E_\text{Auger} = 32\,\text{EeV} \leftrightarrow E_\text{TA} = 40.0\,\mathrm{EeV}$. We have $2\,936$ Auger events above~$32\,\text{EeV}$ and $513$ TA events above~$40.0\,\text{EeV}$ (of which 285 TA strict-cut events with~$40.0\,\text{EeV} \le E_\text{TA} < 57\,\mathrm{EeV}$ and 228 TA loose-cut events with~$E_\text{TA} \ge 57\,\mathrm{EeV}$).

\section{Large-scale analysis}
\label{Sec: Large-Scale}

\begin{figure}
        \centering
        \includegraphics[scale=0.99]{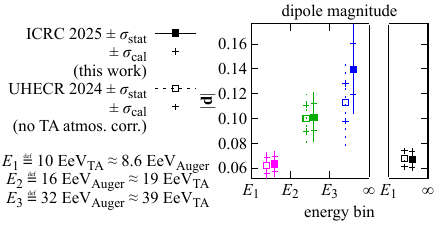}%
        \includegraphics[scale=0.99]{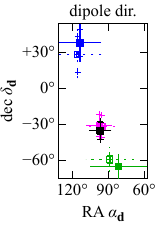}%
        \includegraphics[scale=0.99,page=4]{Figures/aps.pdf}\\
        \includegraphics[scale=0.99,page=1]{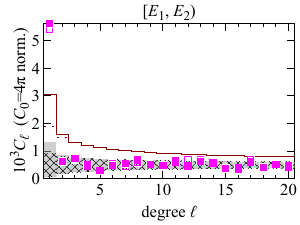}%
        \includegraphics[scale=0.99,page=2]{Figures/aps.pdf}%
        \includegraphics[scale=0.99,page=3]{Figures/aps.pdf}

    \caption{The first and second panels show how the magnitude and direction of the dipole have changed with the inclusion of TA atmospheric corrections; the next ones show how the harmonic space auto-correlation (also known as angular power spectrum) has changed in the different energy bins used.}
    \label{fig:Large-Scale}
\end{figure}

Given our data map of UHECRs, we can expand it in spherical harmonics as
\[M(\textbf{n}) =  \sum_{\ell=0}^{\infty} \sum_{m=-\ell}^{+\ell} a_{\ell m}Y_{\ell m}(\textbf{n})  \to a_{\ell m} = \int d\Omega\, Y_{\ell m}(\textbf{n}) M(\textbf{n}).\]
The $\ell=1$ contribution can be expressed as a dipole vector $\boldsymbol{d} = \sqrt{3}(a_{11},a_{1-1},a_{10})$, its direction and magnitude shown in the first two plots of \autoref{fig:Large-Scale}.
Since we have a full-sky dataset, we can directly measure ~$a_{\ell m} = \sum_\text{evt} Y_{\ell m}(\textbf{n}_\text{evt})/\omega(\textbf{n}_\text{evt})$, where $\omega$~is the total directional exposure of the combined dataset. We perform the analysis in three energy bins $[E_1, E_2)$, $[E_2, E_3)$, $[E_3, +\infty)$ as defined in the previous section plus a cumulative bin $[E_3, +\infty)$. 

The results are shown in \autoref{fig:Large-Scale}, and compared with our previous results \cite{Auger+TA_UHECR2024} using the same time periods but without TA atmospheric corrections. The dipole and quadrupole in the highest energy bin have become stronger, but still not particularly significant, respectively at $p = 0.011$~($2.5\sigma$) and $p = 0.0041$~($2.9\sigma$) pre-trial. All results are otherwise similar to the previous ones.

The dipole direction in the cumulative bin is $92\degree$ away from the Galactic Centre. A hint of a quadrupole along the supergalactic plane is found in the highest energy bin, motivating \autoref{Sec:Harmonic}.

\section{Intermediate-scale analysis}
\label{Sec: Inter-Scale}

This analysis must be performed at the highest energies, where the deflections due to magnetic fields are expected to be smaller. We perform targeted searches using four source catalogues \cite{Auger_MSA}:
\begin{itemize}
    \item 
    44\,113 galaxies of all types, based on the 2MASS catalogue, weighted by their near-IR fluxes;
    \item 
    44 starburst galaxies, based on the Lunardini catalogue, weighted by their radio fluxes;
    \item 
    523 AGNs, based on the \textit{Swift}-BAT catalogue, weighted by their X-ray fluxes; and
    \item 
    26 jetted AGNs, from the \textit{Fermi}-LAT catalogue, weighted by their gamma-ray fluxes.
\end{itemize}
For more details about the data sources, selection criteria, and weights used for compiling these catalogues see \cite{Auger_MSA}.
All these catalogues include galaxies with distances $1\,\mathrm{Mpc} \le D < 250\,\mathrm{Mpc}$, except the starburst galaxy one, which includes galaxies with distances $1\,\mathrm{Mpc} \le D < 130\,\mathrm{Mpc}$.

We then define the test statistics \begin{align*}
    \text{TS}(\Theta, f, E_\text{min}) &= 2\ln\left(\frac{\mathcal{L}(\Theta,f,E_\text{min})}{\mathcal{L}(\Theta,f = 0,E_\text{min})} \right), &
    \mathcal{L}(\Theta,f,E_\text{min}) = \prod_{E_i >E_\text{min}} \frac{\Phi(\boldsymbol{\hat{n}}_i,\Theta,f)\omega(\boldsymbol{\hat{n}}_i,E_i)}{\int_{4\pi}d\Omega\Phi(\boldsymbol{\hat{n}},\Theta,f)\omega(\boldsymbol{\hat{n}},E_i)}.
\end{align*} Here, $i$ refers to the $i$-th UHECR, while the weight $\omega(\boldsymbol{\hat{n}}, E)$ is the directional exposure, which (unlike in our previous studies \cite{Auger_MSA,Auger+TA_UHECR2024}) is also a function of the energy because a TA dataset with stricter cuts is used for $40\,\mathrm{EeV} \le E_\text{TA} < 57\,\mathrm{EeV}$ but one with looser cuts, hence larger exposure, is used for $E_\text{TA} \ge 57\,\mathrm{EeV}$; and $\Phi(\boldsymbol{\hat{n}}_i,\Theta,f)$ is defined as: \[
    \Phi(\boldsymbol{\hat{n}},\Theta,f) := f\Phi_{\text{signal}}(\boldsymbol{\hat{n}},\Theta) + (1-f)\Phi_{\text{background}},
\] where the background contribution is isotropic and the signal contribution is a sum of  von Mises--
Fisher distributions for each source
\begin{align*}
\Phi_{\text{signal}}(\boldsymbol{\hat{n}},\Theta) &\mathrel{:=} \frac{1}{\sum_s a(D_s,E_{\min}) w_s}\sum_s  \frac{a(D_s,E_{\min}) w_s\Theta^{-2}}{4\pi \text{sinh}(\Theta^{-2})}\,\text{e}^{\Theta^{-2}\boldsymbol{\hat{n}}_s \cdot \boldsymbol{\hat{n}}}, &
\Phi_{\text{background}} &\mathrel{:=} \frac{1}{4\pi},
\end{align*}
where the index~$s$ runs over sources in the galaxy catalogue, $w_s$ is the flux of the $s$-th source, and $a(D_s,E_{\min})$ is the attenuation, computed as in \cite{Auger_MSA} following the best fit to Auger data above the ankle energy \cite{Auger_JCAP2017} assuming the \textsc{Epos}~LHC hadronic interaction model, with a mixed mass composition~($67.3\%$~He, $28.1\%$~N and $4.6\%$~Si at $E=1\,\mathrm{EeV}$) and a relatively hard spectrum~($\propto E^{-0.96}\exp(-R/10^{18.68}\,\mathrm{V})$ where~$R=E/Z$) at injection. 

\begin{figure}
    \centering
    \includegraphics{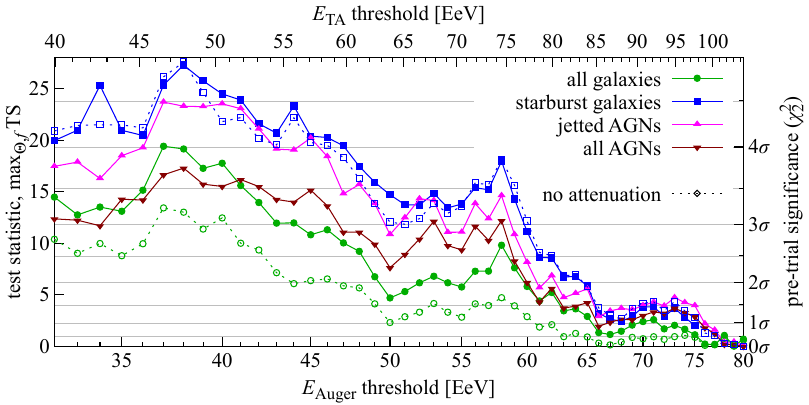}
    \caption{Significances of the correlations with galaxy catalogues as a function of the energy threshold.}
    \label{fig:bestfit}
\end{figure}
The analysis is repeated for increasing energy thresholds, $E_\text{Auger} = 32\,\mathrm{EeV}, 33\,\mathrm{EeV}, \dots, 80 \,\text{EeV}$. The results are presented in \autoref{fig:bestfit}. 
\begin{table}
    \centering
    \renewcommand{~}{\phantom{0}}
    \begin{tblr}{
        vline{1} = {2-7}{},
        vline{2,7} = {-}{},
        hline{1} = {2-6}{},
        hline{2,6,8} = {-}{},
    }
        & $E_\text{min}$ & $\mathrm{TS}$ & $f/\%$ & $\Theta/\degree$ & post-trial \\
        All             galaxies        & $37\,\mathrm{EeV}_\text{Auger} \approx 47\,\mathrm{EeV}_\text{TA}$ & 19.3 & $13.1_{-3.6~}^{+4.7~}$ & $15.5_{-3.6~}^{+6.1~}$ & $3.4\sigma$ \\
        Starburst       galaxies        & $38\,\mathrm{EeV}_\text{Auger} \approx 48\,\mathrm{EeV}_\text{TA}$ & 27.3 & $10.6_{-3.2~}^{+56.6}$ & $17.6_{-4.1~}^{+26.6}$ & $4.2\sigma$ \\
        All             AGNs            & $38\,\mathrm{EeV}_\text{Auger} \approx 48\,\mathrm{EeV}_\text{TA}$ & 17.6 & $~4.8_{-1.4~}^{+1.6~}$ & $15.4_{-2.8~}^{+3.5~}$ & $3.3\sigma$ \\
        Jetted          AGNs            & $37\,\mathrm{EeV}_\text{Auger} \approx 47\,\mathrm{EeV}_\text{TA}$ & 22.9 & $~8.8_{-2.3~}^{-2.6~}$ & $17.4_{-2.8~}^{+3.4~}$ & $3.8\sigma$ \\
        All gal.\ (no atten.)     & $37\,\mathrm{EeV}_\text{Auger} \approx 47\,\mathrm{EeV}_\text{TA}$ & 13.5 & $33.6_{-19.4}^{+26.3}$ & $29.2_{-17.5}^{+12.9}$ & $2.8\sigma$ \\
        Starburst gal.\ (no atten.)  & $38\,\mathrm{EeV}_\text{Auger} \approx 48\,\mathrm{EeV}_\text{TA}$ & 27.3 & $10.6_{-2.7~}^{+4.0~}$ & $15.0_{-2.9~}^{+4.8~}$ & $4.2\sigma$ \\     
    \end{tblr}
    \caption{Best fit of the medium-scale analysis for each of the galaxy catalogues considered. We show both the optimal parameters and the $\mathrm{TS}$ that was achieved, as well as the corresponding one-tailed post-trial significance. The reasons for the increase in the upper uncertainties on $f$ and $\Theta$ when including the attenuation in the starburst galaxy catalogue are under investigation.}
    \label{tab:Int-Anis}
\end{table}
The details of the maximum TS for each galaxy catalogue are shown in \autoref{tab:Int-Anis}.
To allow a direct comparison to our previous results \cite{Auger+TA_UHECR2024} where the TA loose-cut data were not used, in the case of the all-galaxy and starburst galaxy catalogues we also shown results obtained neglecting the attenuation.

\section{Angular Harmonic Space Analysis}
\label{Sec:Harmonic}

For this analysis we combine the Harmonic Auto-Correlation as calculated in \autoref{Sec: Large-Scale} with the energy scanning from \autoref{Sec: Inter-Scale}. We also perform the same analysis with Harmonic Cross-Correlation of the UHECR map and the galaxy catalogues introduced in the previous section (including the attenuations). The Cross-Correlation is a valuable tool as it inherently involves more information than the Auto-Correlation, making it more likely to break possible degeneracies. Furthermore, for the Cross-Correlation the statistical and systematic noises do not correlate, resulting in a clearer signal. In the case in which the number of sources in the catalogue is much bigger than the number of cosmic rays (like is the case for the all-galaxy catalogue), the shot noise is greatly reduced, for more details see \cite{Urban:2020szk}.

Both the Harmonic Auto-Correlation of the UHECRs and their Cross-Correlation with the different galaxy catalogues can be obtained for any multipole following \autoref{Sec: Large-Scale} and using \begin{align*}
    C^{\text{CR CR}}_l &:= \frac{1}{2\ell+1} \sum_{m=-\ell}^{+\ell} a^{\text{CR}}_{lm} a_{lm}^{*\text{CR}}; &
    C^{\text{CR Cat}}_l &:= \frac{1}{2\ell+1} \sum_{m=-\ell}^{+\ell} a^{\text{CR}}_{lm} a_{lm}^{*\text{Cat}}.
\end{align*}

For the Cross-Correlation, the map used includes the attenuation discussed in \autoref{Sec: Inter-Scale}. Then, using isotropic simulations, we can obtain the pre-trial significance of each multipole (we consider $\ell_\text{max} = 20$ in this work). To reduce computational time, it is assumed that isotropic Cross-Correlations follow a Gaussian distribution, while the Auto-Correlations follow \cite{Percival:2006ss}: \[
    f(\bar{C}_\ell^\text{CR CR}|C^{\text{CR CR}}_\ell) \propto C_\ell^\text{CR CR} \left(\frac{\bar{C}_\ell^\text{CR CR}}{C_\ell^\text{CR CR}} \right)^{\frac{2\ell + 1}{2}-1} \text{exp}\left( -\frac{(2\ell + 1)\bar{C}_\ell^\text{CR CR}}{2}C_\ell^\text{CR CR} \right),
\] where $\bar{C}_\ell^\text{CR CR}$ is expected value for isotropy, given by the average of all simulations. We can then calculate the $p$-value of a data point $C_{\ell, \text{Data}}^\text{CR CR}$ as \[
    \text{$p$-value} = 1-\frac{\int_0^{C_{\ell, \text{Data}}^\text{CR CR}} dC^{\text{CR CR}}_\ell \, f(\bar{C}_\ell^\text{CR CR}|C^{\text{CR CR}}_\ell)}{\int_0^{\infty}dC^{\text{CR CR}}_\ell \, f(\bar{C}_\ell^\text{CR CR}|C^{\text{CR CR}}_\ell)},
\]
and equivalently for the cross-correlation, for which the function $f$ would be a Gaussian with  mean and standard deviation calculated directly from the isotropic simulations. Then the analysis is repeated for increasing energy thresholds as in \autoref{Sec: Inter-Scale}. For the Cross-Correlations we use the same galaxy catalogues as in  \autoref{Sec: Inter-Scale}.

\begin{figure}
			\centering
            \begin{subfigure}{0.32\textwidth}
				\centering
				\includegraphics[width=\textwidth]{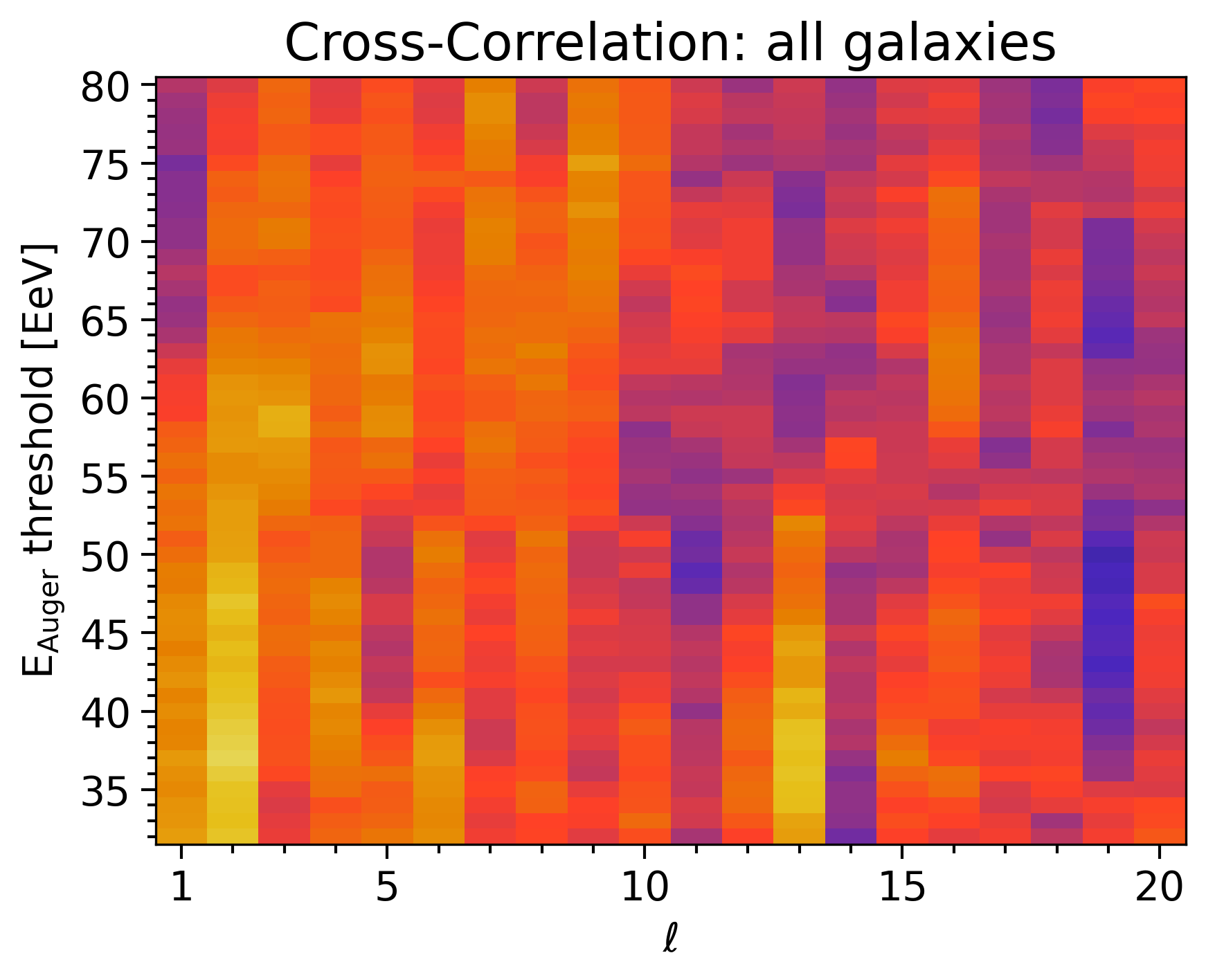}
			\end{subfigure}
			\begin{subfigure}{0.32\textwidth}
				\centering
				\includegraphics[width=\textwidth]{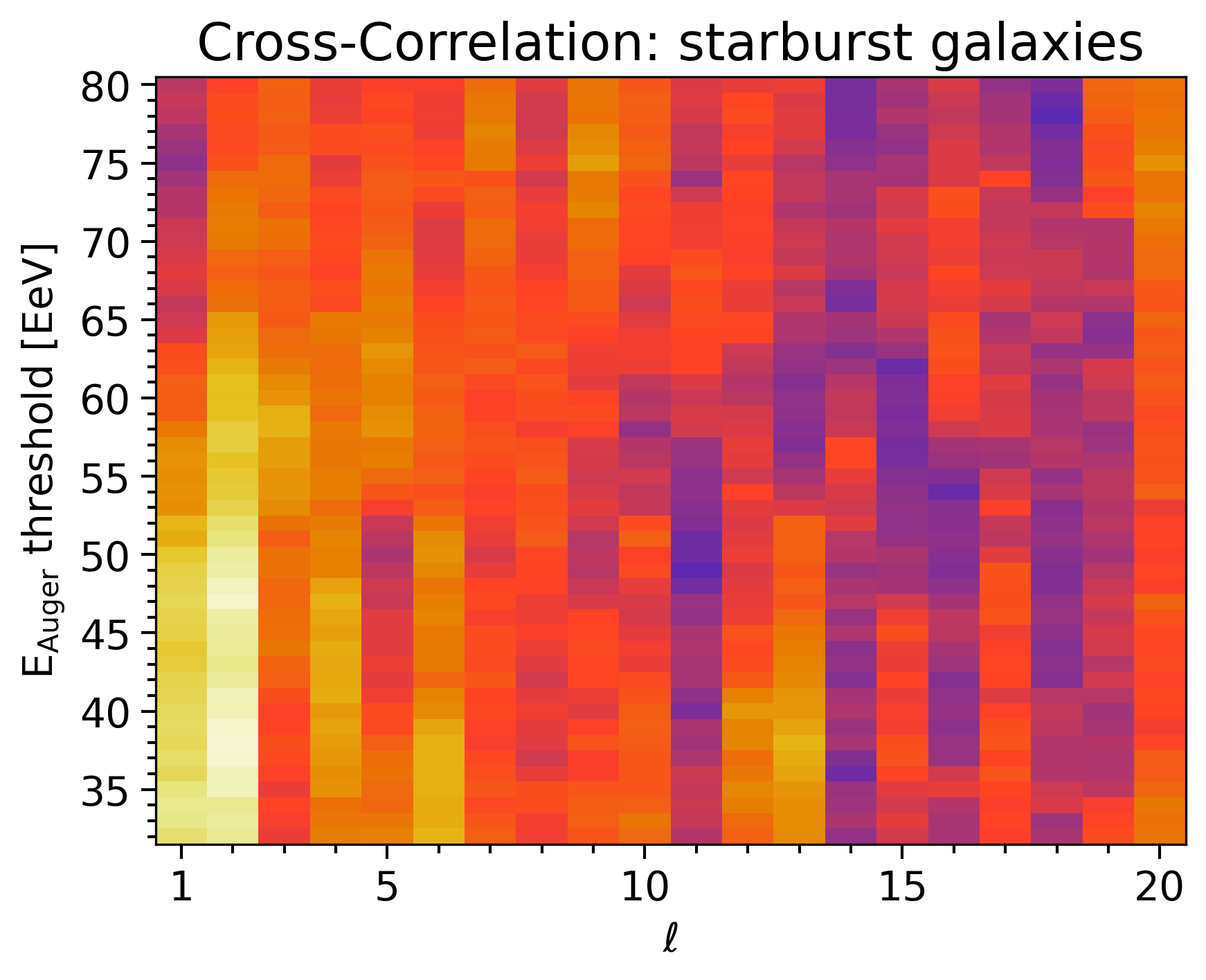}
			\end{subfigure}
			\begin{subfigure}{0.32\textwidth}
				\centering
				\includegraphics[width=\textwidth]{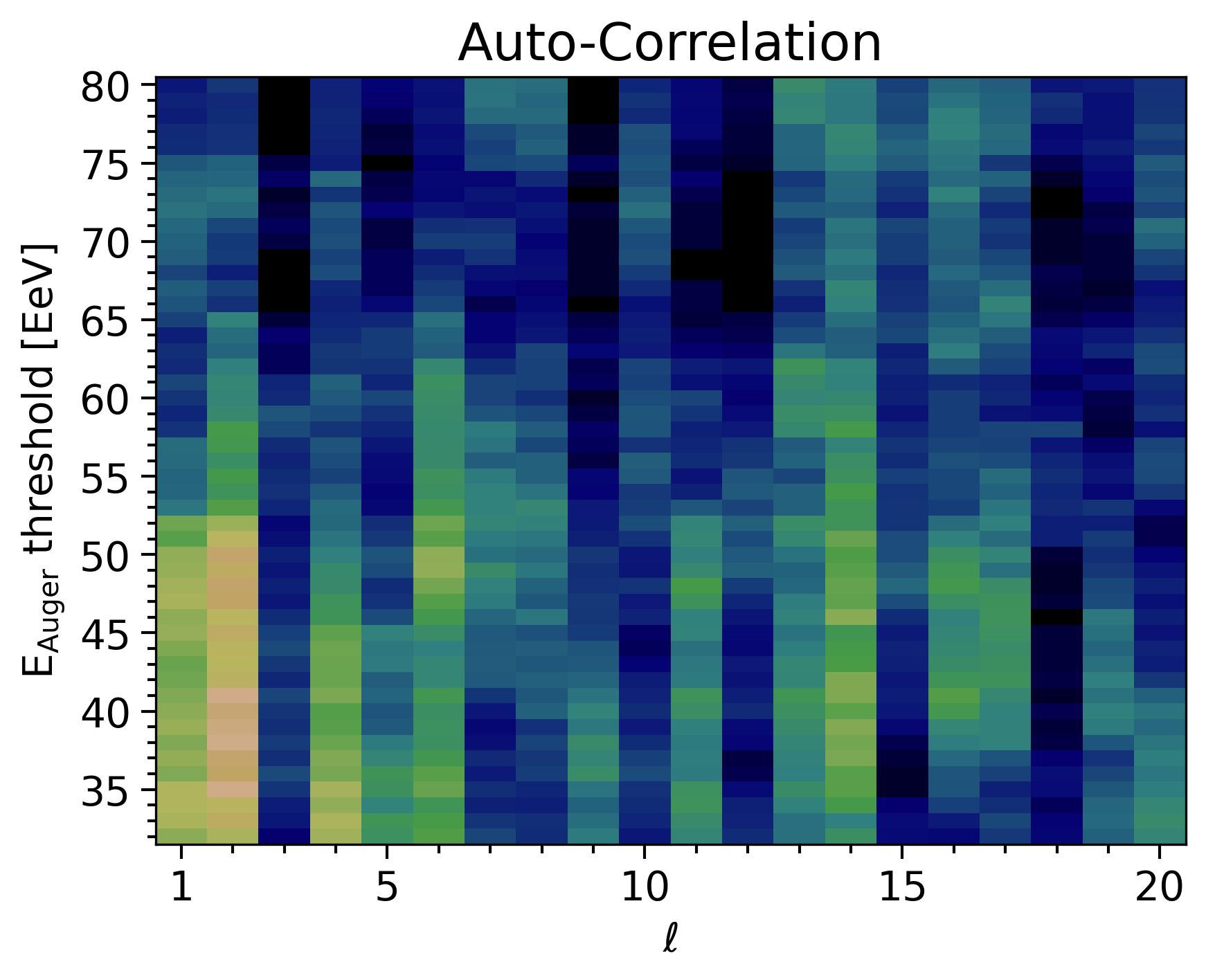}
			\end{subfigure}
			\hfill
            \begin{subfigure}{0.32\textwidth}
				\centering
				\includegraphics[width=\textwidth]{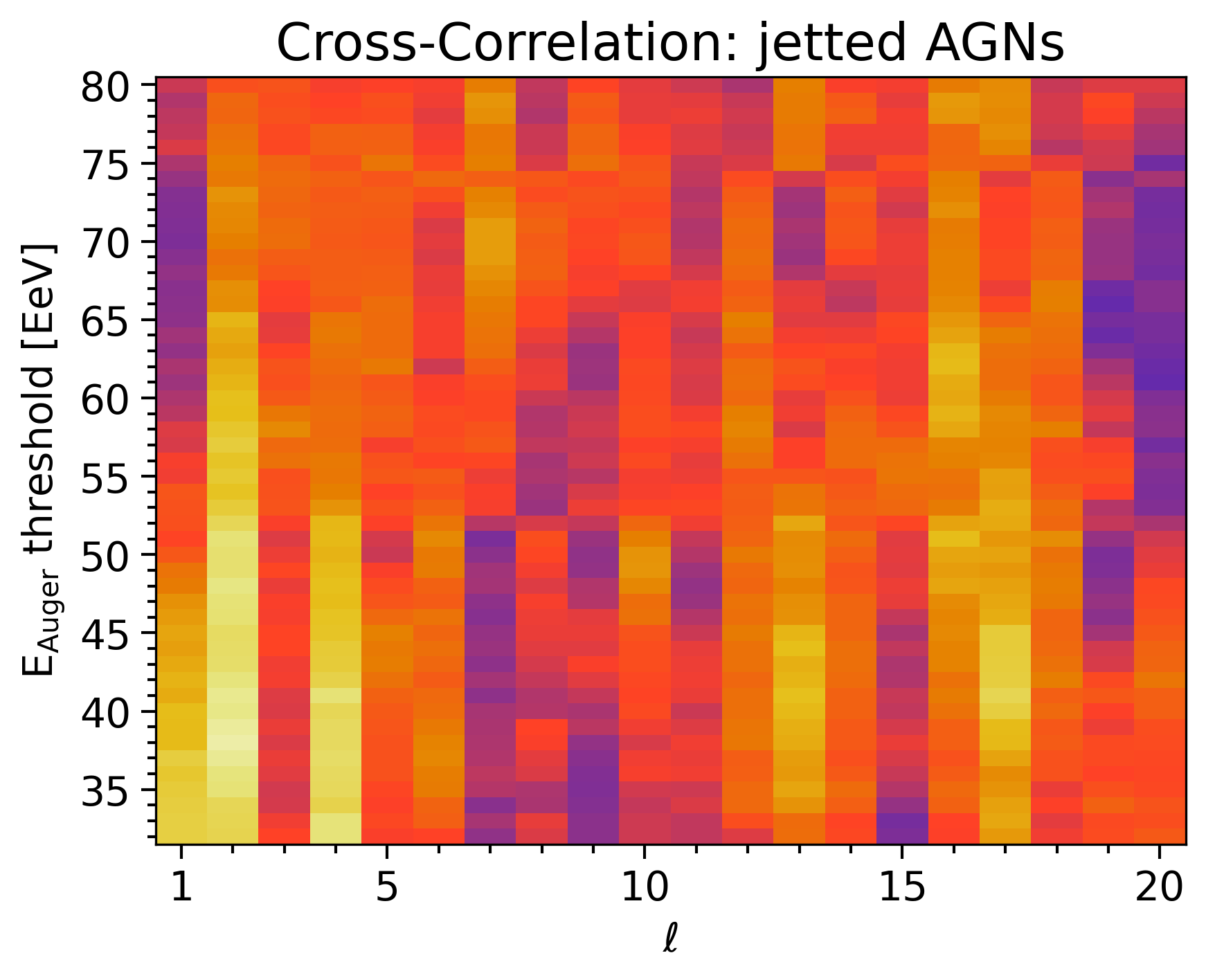}
			\end{subfigure}
			\begin{subfigure}{0.32\textwidth}
				\centering
				\includegraphics[width=\textwidth]{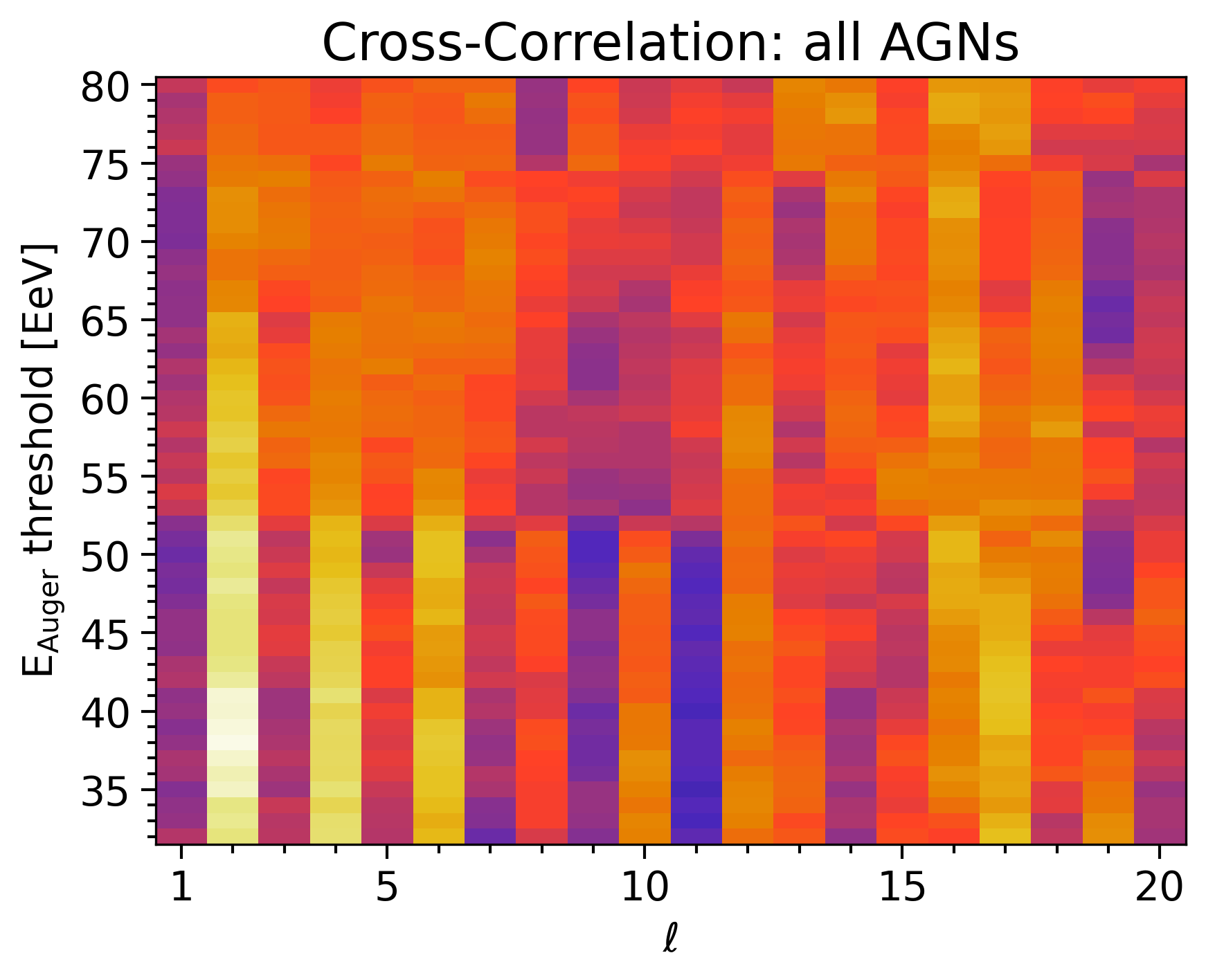}
			\end{subfigure}
            \begin{subfigure}{0.32\textwidth}
				\centering
				\includegraphics[width=\textwidth, height=4cm, keepaspectratio]{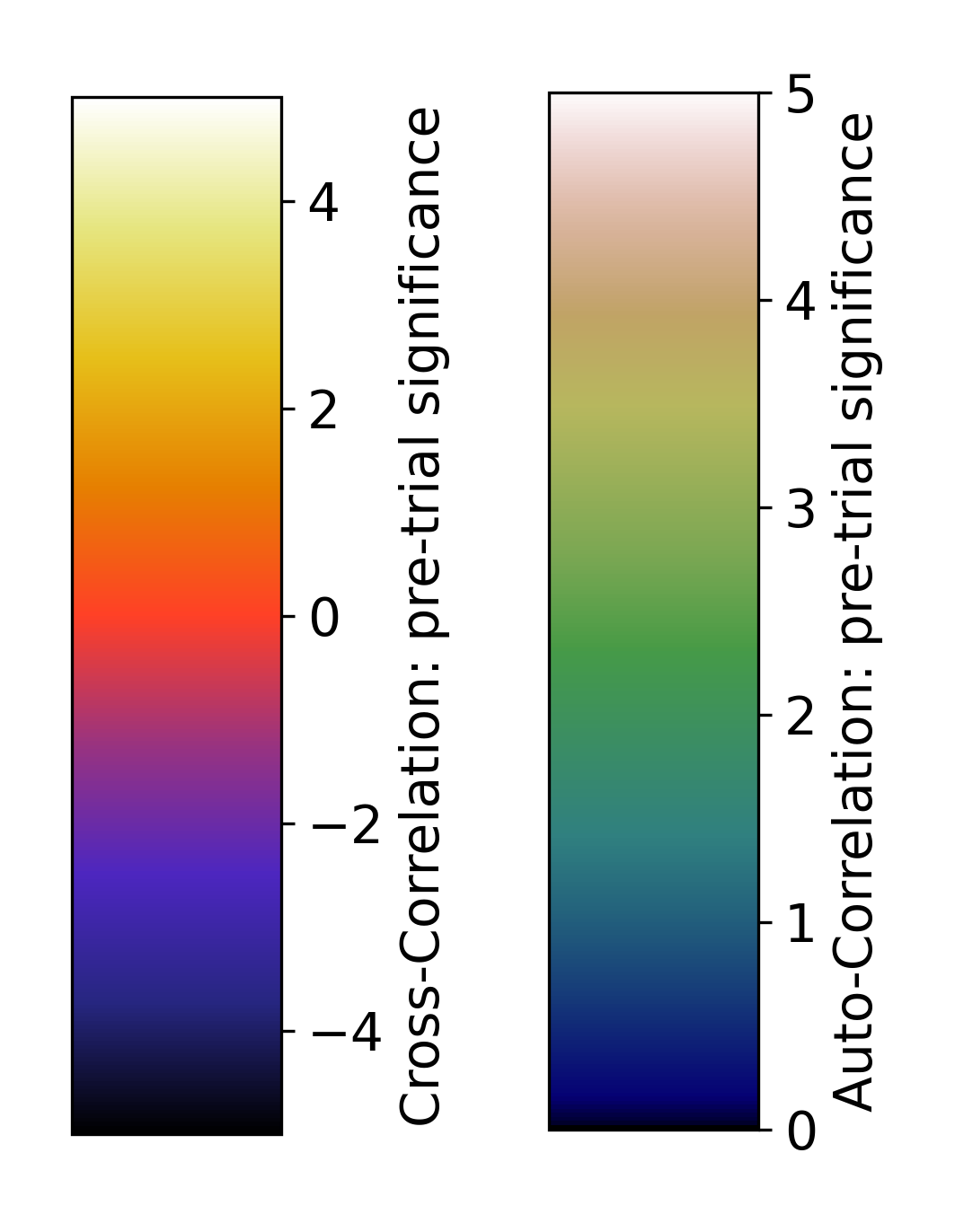}
			\end{subfigure}
			\hfill
            \caption{Pre-trial significances for each multipole order and energy threshold of the Auto-Correlation and the Cross-Correlation with all galaxies, starburst galaxies, jetted AGNs  and all AGNs.}
            \label{fig:Harmonic}
\end{figure}

The results are shown in \autoref{fig:Harmonic}. From this figure we see that the quadrupole is the most statistically significant multipole in all cases. We then find the energy threshold at which the quadrupole is the most significant in each plot of \autoref{fig:Harmonic} and obtain a corresponding post-trial significance. For this we follow the most conservative approach, by taking into account the scan in energy and the measurements of different mutipoles up to $ \ell = 20$. The results for this analysis are then presented in \autoref{tab:Harmonic}.

\begin{table}
\centering
\begin{tblr}{
  vline{1-2,6} = {-}{},
  hline{1-2,7} = {-}{},
}
                     & $\ell$ & $E_{\min}$ & pre-trial & post-trial \\
Auto-Correlation     &    2    &     $41\,\mathrm{EeV}_\text{Auger} \approx 51.8\,\mathrm{EeV}_\text{TA}$    &4.2$\sigma$&2.1$\sigma$   \\
All galaxies                &    2    &     $37\,\mathrm{EeV}_\text{Auger} \approx 46.5\,\mathrm{EeV}_\text{TA}$    &3.2$\sigma$&      ---     \\
Starburst galaxies            &    2    &     $38\,\mathrm{EeV}_\text{Auger} \approx 47.8\,\mathrm{EeV}_\text{TA}$     &4.5$\sigma$&2.7$\sigma$   \\
All AGNs            &    2    &     $38\,\mathrm{EeV}_\text{Auger} \approx 47.8\,\mathrm{EeV}_\text{TA}$     &4.7$\sigma$&3.0$\sigma$   \\
Jetted AGNs            &    2    &     $38\,\mathrm{EeV}_\text{Auger} \approx 47.8\,\mathrm{EeV}_\text{TA}$     &4.1$\sigma$&2.0$\sigma$    
\end{tblr}
\caption{We show the $\ell$ and $E_{\min}$ at which the maximum significance is obtained for the Auto-Correlation and for the Cross-Correlation with each galaxy catalogue. We present the pre-trial maximum obtained and the post-trial result calculated from it.}
\label{tab:Harmonic}
\end{table}

\section{Conclusions}

We have presented an update to our previous analyses and introduced a new one. For the analysis on large angular scales, we have incorporated new atmospheric corrections to the energy of the TA data. Most of our results remain similar, except for those at the highest energies, where the dipole and quadrupole have become more significant (but within the statistical uncertainties).

For our intermediate-scale analysis, we have introduced new data from the TA side, using a dataset with looser cuts at the highest energies. For the first time in full-sky analyses, we have also taken into account the energy losses of UHECRs. Furthermore, we have extended the analysis to include two more source catalogues. The correlation with starburst galaxies remains the most significant, at $4.2\sigma$ post-trial, even after the inclusion of attenuations, which primarily enhanced the correlation with the all-galaxy catalogue.

Finally, we have studied the harmonic autocorrelation and cross-correlation methods. Our results show that, given an appropriate source catalogue, the cross-correlation can yield a clearer signal than the harmonic auto-correlation. As a next step, we plan to investigate how to combine auto- and cross-correlations, or how to merge different multipoles to enhance statistical significance. In the future, the method introduced here could serve as a robust tool for source matching, less sensitive to the specific characteristics of the Galactic magnetic field.

\pagebreak
\section*{The Pierre Auger Collaboration}

{\footnotesize\setlength{\baselineskip}{10pt}
\noindent
\begin{wrapfigure}[11]{l}{0.12\linewidth}
\vspace{-4pt}
\includegraphics[width=0.98\linewidth]{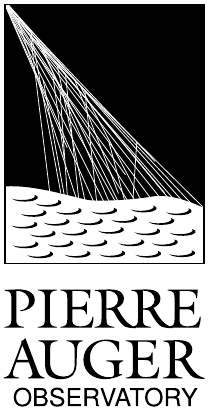}
\end{wrapfigure}
\begin{sloppypar}\noindent
A.~Abdul Halim$^{13}$,
P.~Abreu$^{70}$,
M.~Aglietta$^{53,51}$,
I.~Allekotte$^{1}$,
K.~Almeida Cheminant$^{78,77}$,
A.~Almela$^{7,12}$,
R.~Aloisio$^{44,45}$,
J.~Alvarez-Mu\~niz$^{76}$,
A.~Ambrosone$^{44}$,
J.~Ammerman Yebra$^{76}$
% ...

\end{sloppypar}
\begin{center}
\end{center}

\vspace{1ex}
\begin{description}[labelsep=0.2em,align=right,labelwidth=0.7em,labelindent=0em,leftmargin=2em,noitemsep,before={\renewcommand\makelabel[1]{##1 }}]
\item[$^{1}$] Centro At\'omico Bariloche and Instituto Balseiro (CNEA-UNCuyo-CONICET), San Carlos de Bariloche, Argentina
\item[$^{2}$] Departamento de F\'\i{}sica and Departamento de Ciencias de la Atm\'osfera y los Oc\'eanos, FCEyN, Universidad de Buenos Aires and CONICET, Buenos Aires, Argentina
\item[$^{3}$] IFLP, Universidad Nacional de La Plata and CONICET, La Plata, Argentina
\item[$^{4}$] Instituto de Astronom\'\i{}a y F\'\i{}sica del Espacio (IAFE, CONICET-UBA), Buenos Aires, Argentina
\item[$^{5}$] Instituto de F\'\i{}sica de Rosario (IFIR) -- CONICET/U.N.R.\ and Facultad de Ciencias Bioqu\'\i{}micas y Farmac\'euticas U.N.R., Rosario, Argentina
\item[$^{6}$] Instituto de Tecnolog\'\i{}as en Detecci\'on y Astropart\'\i{}culas (CNEA, CONICET, UNSAM), and Universidad Tecnol\'ogica Nacional -- Facultad Regional Mendoza (CONICET/CNEA), Mendoza, Argentina
\item[$^{7}$] Instituto de Tecnolog\'\i{}as en Detecci\'on y Astropart\'\i{}culas (CNEA, CONICET, UNSAM), Buenos Aires, Argentina
\item[$^{8}$] International Center of Advanced Studies and Instituto de Ciencias F\'\i{}sicas, ECyT-UNSAM and CONICET, Campus Miguelete -- San Mart\'\i{}n, Buenos Aires, Argentina
\item[$^{9}$] Laboratorio Atm\'osfera -- Departamento de Investigaciones en L\'aseres y sus Aplicaciones -- UNIDEF (CITEDEF-CONICET), Argentina
\item[$^{10}$] Observatorio Pierre Auger, Malarg\"ue, Argentina
\item[$^{11}$] Observatorio Pierre Auger and Comisi\'on Nacional de Energ\'\i{}a At\'omica, Malarg\"ue, Argentina
\item[$^{12}$] Universidad Tecnol\'ogica Nacional -- Facultad Regional Buenos Aires, Buenos Aires, Argentina
\item[$^{13}$] University of Adelaide, Adelaide, S.A., Australia
\item[$^{14}$] Universit\'e Libre de Bruxelles (ULB), Brussels, Belgium
\item[$^{15}$] Vrije Universiteit Brussels, Brussels, Belgium
\item[$^{16}$] Centro Brasileiro de Pesquisas Fisicas, Rio de Janeiro, RJ, Brazil
\item[$^{17}$] Centro Federal de Educa\c{c}\~ao Tecnol\'ogica Celso Suckow da Fonseca, Petropolis, Brazil
\item[$^{18}$] Instituto Federal de Educa\c{c}\~ao, Ci\^encia e Tecnologia do Rio de Janeiro (IFRJ), Brazil
\item[$^{19}$] Universidade de S\~ao Paulo, Escola de Engenharia de Lorena, Lorena, SP, Brazil
\item[$^{20}$] Universidade de S\~ao Paulo, Instituto de F\'\i{}sica de S\~ao Carlos, S\~ao Carlos, SP, Brazil
\item[$^{21}$] Universidade de S\~ao Paulo, Instituto de F\'\i{}sica, S\~ao Paulo, SP, Brazil
\item[$^{22}$] Universidade Estadual de Campinas (UNICAMP), IFGW, Campinas, SP, Brazil
\item[$^{23}$] Universidade Estadual de Feira de Santana, Feira de Santana, Brazil
\item[$^{24}$] Universidade Federal de Campina Grande, Centro de Ciencias e Tecnologia, Campina Grande, Brazil
\item[$^{25}$] Universidade Federal do ABC, Santo Andr\'e, SP, Brazil
\item[$^{26}$] Universidade Federal do Paran\'a, Setor Palotina, Palotina, Brazil
\item[$^{27}$] Universidade Federal do Rio de Janeiro, Instituto de F\'\i{}sica, Rio de Janeiro, RJ, Brazil
\item[$^{28}$] Universidad de Medell\'\i{}n, Medell\'\i{}n, Colombia
\item[$^{29}$] Universidad Industrial de Santander, Bucaramanga, Colombia
\item[$^{30}$] Charles University, Faculty of Mathematics and Physics, Institute of Particle and Nuclear Physics, Prague, Czech Republic
\item[$^{31}$] Institute of Physics of the Czech Academy of Sciences, Prague, Czech Republic
\item[$^{32}$] Palacky University, Olomouc, Czech Republic
\item[$^{33}$] CNRS/IN2P3, IJCLab, Universit\'e Paris-Saclay, Orsay, France
\item[$^{34}$] Laboratoire de Physique Nucl\'eaire et de Hautes Energies (LPNHE), Sorbonne Universit\'e, Universit\'e de Paris, CNRS-IN2P3, Paris, France
\item[$^{35}$] Univ.\ Grenoble Alpes, CNRS, Grenoble Institute of Engineering Univ.\ Grenoble Alpes, LPSC-IN2P3, 38000 Grenoble, France
\item[$^{36}$] Universit\'e Paris-Saclay, CNRS/IN2P3, IJCLab, Orsay, France
\item[$^{37}$] Bergische Universit\"at Wuppertal, Department of Physics, Wuppertal, Germany
\item[$^{38}$] Karlsruhe Institute of Technology (KIT), Institute for Experimental Particle Physics, Karlsruhe, Germany
\item[$^{39}$] Karlsruhe Institute of Technology (KIT), Institut f\"ur Prozessdatenverarbeitung und Elektronik, Karlsruhe, Germany
\item[$^{40}$] Karlsruhe Institute of Technology (KIT), Institute for Astroparticle Physics, Karlsruhe, Germany
\item[$^{41}$] RWTH Aachen University, III.\ Physikalisches Institut A, Aachen, Germany
\item[$^{42}$] Universit\"at Hamburg, II.\ Institut f\"ur Theoretische Physik, Hamburg, Germany
\item[$^{43}$] Universit\"at Siegen, Department Physik -- Experimentelle Teilchenphysik, Siegen, Germany
\item[$^{44}$] Gran Sasso Science Institute, L'Aquila, Italy
\item[$^{45}$] INFN Laboratori Nazionali del Gran Sasso, Assergi (L'Aquila), Italy
\item[$^{46}$] INFN, Sezione di Catania, Catania, Italy
\item[$^{47}$] INFN, Sezione di Lecce, Lecce, Italy
\item[$^{48}$] INFN, Sezione di Milano, Milano, Italy
\item[$^{49}$] INFN, Sezione di Napoli, Napoli, Italy
\item[$^{50}$] INFN, Sezione di Roma ``Tor Vergata'', Roma, Italy
\item[$^{51}$] INFN, Sezione di Torino, Torino, Italy
\item[$^{52}$] Istituto di Astrofisica Spaziale e Fisica Cosmica di Palermo (INAF), Palermo, Italy
\item[$^{53}$] Osservatorio Astrofisico di Torino (INAF), Torino, Italy
\item[$^{54}$] Politecnico di Milano, Dipartimento di Scienze e Tecnologie Aerospaziali , Milano, Italy
\item[$^{55}$] Universit\`a del Salento, Dipartimento di Matematica e Fisica ``E.\ De Giorgi'', Lecce, Italy
\item[$^{56}$] Universit\`a dell'Aquila, Dipartimento di Scienze Fisiche e Chimiche, L'Aquila, Italy
\item[$^{57}$] Universit\`a di Catania, Dipartimento di Fisica e Astronomia ``Ettore Majorana``, Catania, Italy
\item[$^{58}$] Universit\`a di Milano, Dipartimento di Fisica, Milano, Italy
\item[$^{59}$] Universit\`a di Napoli ``Federico II'', Dipartimento di Fisica ``Ettore Pancini'', Napoli, Italy
\item[$^{60}$] Universit\`a di Palermo, Dipartimento di Fisica e Chimica ''E.\ Segr\`e'', Palermo, Italy
\item[$^{61}$] Universit\`a di Roma ``Tor Vergata'', Dipartimento di Fisica, Roma, Italy
\item[$^{62}$] Universit\`a Torino, Dipartimento di Fisica, Torino, Italy
\item[$^{63}$] Benem\'erita Universidad Aut\'onoma de Puebla, Puebla, M\'exico
\item[$^{64}$] Unidad Profesional Interdisciplinaria en Ingenier\'\i{}a y Tecnolog\'\i{}as Avanzadas del Instituto Polit\'ecnico Nacional (UPIITA-IPN), M\'exico, D.F., M\'exico
\item[$^{65}$] Universidad Aut\'onoma de Chiapas, Tuxtla Guti\'errez, Chiapas, M\'exico
\item[$^{66}$] Universidad Michoacana de San Nicol\'as de Hidalgo, Morelia, Michoac\'an, M\'exico
\item[$^{67}$] Universidad Nacional Aut\'onoma de M\'exico, M\'exico, D.F., M\'exico
\item[$^{68}$] Institute of Nuclear Physics PAN, Krakow, Poland
\item[$^{69}$] University of \L{}\'od\'z, Faculty of High-Energy Astrophysics,\L{}\'od\'z, Poland
\item[$^{70}$] Laborat\'orio de Instrumenta\c{c}\~ao e F\'\i{}sica Experimental de Part\'\i{}culas -- LIP and Instituto Superior T\'ecnico -- IST, Universidade de Lisboa -- UL, Lisboa, Portugal
\item[$^{71}$] ``Horia Hulubei'' National Institute for Physics and Nuclear Engineering, Bucharest-Magurele, Romania
\item[$^{72}$] Institute of Space Science, Bucharest-Magurele, Romania
\item[$^{73}$] Center for Astrophysics and Cosmology (CAC), University of Nova Gorica, Nova Gorica, Slovenia
\item[$^{74}$] Experimental Particle Physics Department, J.\ Stefan Institute, Ljubljana, Slovenia
\item[$^{75}$] Universidad de Granada and C.A.F.P.E., Granada, Spain
\item[$^{76}$] Instituto Galego de F\'\i{}sica de Altas Enerx\'\i{}as (IGFAE), Universidade de Santiago de Compostela, Santiago de Compostela, Spain
\item[$^{77}$] IMAPP, Radboud University Nijmegen, Nijmegen, The Netherlands
\item[$^{78}$] Nationaal Instituut voor Kernfysica en Hoge Energie Fysica (NIKHEF), Science Park, Amsterdam, The Netherlands
\item[$^{79}$] Stichting Astronomisch Onderzoek in Nederland (ASTRON), Dwingeloo, The Netherlands
\item[$^{80}$] Universiteit van Amsterdam, Faculty of Science, Amsterdam, The Netherlands
\item[$^{81}$] Case Western Reserve University, Cleveland, OH, USA
\item[$^{82}$] Colorado School of Mines, Golden, CO, USA
\item[$^{83}$] Department of Physics and Astronomy, Lehman College, City University of New York, Bronx, NY, USA
\item[$^{84}$] Michigan Technological University, Houghton, MI, USA
\item[$^{85}$] New York University, New York, NY, USA
\item[$^{86}$] University of Chicago, Enrico Fermi Institute, Chicago, IL, USA
\item[$^{87}$] University of Delaware, Department of Physics and Astronomy, Bartol Research Institute, Newark, DE, USA
\item[] -----
\item[$^{a}$] Max-Planck-Institut f\"ur Radioastronomie, Bonn, Germany
\item[$^{b}$] also at Kapteyn Institute, University of Groningen, Groningen, The Netherlands
\item[$^{c}$] School of Physics and Astronomy, University of Leeds, Leeds, United Kingdom
\item[$^{d}$] Fermi National Accelerator Laboratory, Fermilab, Batavia, IL, USA
\item[$^{e}$] Pennsylvania State University, University Park, PA, USA
\item[$^{f}$] Colorado State University, Fort Collins, CO, USA
\item[$^{g}$] Louisiana State University, Baton Rouge, LA, USA
\item[$^{h}$] now at Graduate School of Science, Osaka Metropolitan University, Osaka, Japan
\item[$^{i}$] Institut universitaire de France (IUF), France
\item[$^{j}$] now at Technische Universit\"at Dortmund and Ruhr-Universit\"at Bochum, Dortmund and Bochum, Germany
\end{description}

\section*{Acknowledgments}

\begin{sloppypar}
The successful installation, commissioning, and operation of the Pierre
Auger Observatory would not have been possible without the strong
commitment and effort from the technical and administrative staff in
Malarg\"ue. We are very grateful to the following agencies and
organizations for financial support:
\end{sloppypar}

\begin{sloppypar}
Argentina -- Comisi\'on Nacional de Energ\'\i{}a At\'omica; Agencia Nacional de
Promoci\'on Cient\'\i{}fica y Tecnol\'ogica (ANPCyT); Consejo Nacional de
Investigaciones Cient\'\i{}ficas y T\'ecnicas (CONICET); Gobierno de la
Provincia de Mendoza; Municipalidad de Malarg\"ue; NDM Holdings and Valle
Las Le\~nas; in gratitude for their continuing cooperation over land
access; Australia -- the Australian Research Council; Belgium -- Fonds
de la Recherche Scientifique (FNRS); Research Foundation Flanders (FWO),
Marie Curie Action of the European Union Grant No.~101107047; Brazil --
Conselho Nacional de Desenvolvimento Cient\'\i{}fico e Tecnol\'ogico (CNPq);
Financiadora de Estudos e Projetos (FINEP); Funda\c{c}\~ao de Amparo \`a
Pesquisa do Estado de Rio de Janeiro (FAPERJ); S\~ao Paulo Research
Foundation (FAPESP) Grants No.~2019/10151-2, No.~2010/07359-6 and
No.~1999/05404-3; Minist\'erio da Ci\^encia, Tecnologia, Inova\c{c}\~oes e
Comunica\c{c}\~oes (MCTIC); Czech Republic -- GACR 24-13049S, CAS LQ100102401,
MEYS LM2023032, CZ.02.1.01/0.0/0.0/16{\textunderscore}013/0001402,
CZ.02.1.01/0.0/0.0/18{\textunderscore}046/0016010 and
CZ.02.1.01/0.0/0.0/17{\textunderscore}049/0008422 and CZ.02.01.01/00/22{\textunderscore}008/0004632;
France -- Centre de Calcul IN2P3/CNRS; Centre National de la Recherche
Scientifique (CNRS); Conseil R\'egional Ile-de-France; D\'epartement
Physique Nucl\'eaire et Corpusculaire (PNC-IN2P3/CNRS); D\'epartement
Sciences de l'Univers (SDU-INSU/CNRS); Institut Lagrange de Paris (ILP)
Grant No.~LABEX ANR-10-LABX-63 within the Investissements d'Avenir
Programme Grant No.~ANR-11-IDEX-0004-02; Germany -- Bundesministerium
f\"ur Bildung und Forschung (BMBF); Deutsche Forschungsgemeinschaft (DFG);
Finanzministerium Baden-W\"urttemberg; Helmholtz Alliance for
Astroparticle Physics (HAP); Helmholtz-Gemeinschaft Deutscher
Forschungszentren (HGF); Ministerium f\"ur Kultur und Wissenschaft des
Landes Nordrhein-Westfalen; Ministerium f\"ur Wissenschaft, Forschung und
Kunst des Landes Baden-W\"urttemberg; Italy -- Istituto Nazionale di
Fisica Nucleare (INFN); Istituto Nazionale di Astrofisica (INAF);
Ministero dell'Universit\`a e della Ricerca (MUR); CETEMPS Center of
Excellence; Ministero degli Affari Esteri (MAE), ICSC Centro Nazionale
di Ricerca in High Performance Computing, Big Data and Quantum
Computing, funded by European Union NextGenerationEU, reference code
CN{\textunderscore}00000013; M\'exico -- Consejo Nacional de Ciencia y Tecnolog\'\i{}a
(CONACYT) No.~167733; Universidad Nacional Aut\'onoma de M\'exico (UNAM);
PAPIIT DGAPA-UNAM; The Netherlands -- Ministry of Education, Culture and
Science; Netherlands Organisation for Scientific Research (NWO); Dutch
national e-infrastructure with the support of SURF Cooperative; Poland
-- Ministry of Education and Science, grants No.~DIR/WK/2018/11 and
2022/WK/12; National Science Centre, grants No.~2016/22/M/ST9/00198,
2016/23/B/ST9/01635, 2020/39/B/ST9/01398, and 2022/45/B/ST9/02163;
Portugal -- Portuguese national funds and FEDER funds within Programa
Operacional Factores de Competitividade through Funda\c{c}\~ao para a Ci\^encia
e a Tecnologia (COMPETE); Romania -- Ministry of Research, Innovation
and Digitization, CNCS-UEFISCDI, contract no.~30N/2023 under Romanian
National Core Program LAPLAS VII, grant no.~PN 23 21 01 02 and project
number PN-III-P1-1.1-TE-2021-0924/TE57/2022, within PNCDI III; Slovenia
-- Slovenian Research Agency, grants P1-0031, P1-0385, I0-0033, N1-0111;
Spain -- Ministerio de Ciencia e Innovaci\'on/Agencia Estatal de
Investigaci\'on (PID2019-105544GB-I00, PID2022-140510NB-I00 and
RYC2019-027017-I), Xunta de Galicia (CIGUS Network of Research Centers,
Consolidaci\'on 2021 GRC GI-2033, ED431C-2021/22 and ED431F-2022/15),
Junta de Andaluc\'\i{}a (SOMM17/6104/UGR and P18-FR-4314), and the European
Union (Marie Sklodowska-Curie 101065027 and ERDF); USA -- Department of
Energy, Contracts No.~DE-AC02-07CH11359, No.~DE-FR02-04ER41300,
No.~DE-FG02-99ER41107 and No.~DE-SC0011689; National Science Foundation,
Grant No.~0450696, and NSF-2013199; The Grainger Foundation; Marie
Curie-IRSES/EPLANET; European Particle Physics Latin American Network;
and UNESCO.
\end{sloppypar}

}

\pagebreak

\section*{The Telescope Array Collaboration}

{\footnotesize\setlength{\baselineskip}{10pt}
\noindent
\begin{wrapfigure}{l}{0.15\linewidth}
\vspace{-10pt}
\includegraphics[width=0.98\linewidth]{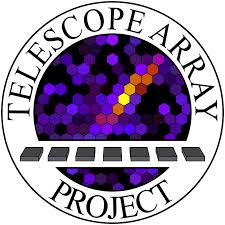}
\end{wrapfigure}
\vspace{-10pt}
\begin{sloppypar}\noindent
R.U.~Abbasi$^{1}$,
T.~Abu-Zayyad$^{1,2}$,
M.~Allen$^{2}$,
J.W.~Belz$^{2}$,
D.R.~Bergman$^{2}$,
F.~Bradfield$^{3}$,
I.~Buckland$^{2}$,
W.~Campbell$^{2}$,
B.G.~Cheon$^{4}$,
K.~Endo$^{3}$
% ...
\bigskip
\par\noindent
{\footnotesize\it
$^{1}$ Department of Physics, Loyola University-Chicago, Chicago, Illinois 60660, USA \\
$^{2}$ High Energy Astrophysics Institute and Department of Physics and Astronomy, University of Utah, Salt Lake City, Utah 84112-0830, USA \\
$^{3}$ Graduate School of Science, Osaka Metropolitan University, Sugimoto, Sumiyoshi, Osaka 558-8585, Japan \\
$^{4}$ Department of Physics and The Research Institute of Natural Science, Hanyang University, Seongdong-gu, Seoul 426-791, Korea \\
% ...

\let\thefootnote\relax\footnote{$\ssymbol{1}$ Presently at: Purple Mountain Observatory, Nanjing 210023, China}
\let\thefootnote\relax\footnote{$\ssymbol{2}$ Presently at: Physics Department, Brookhaven National Laboratory, Upton, NY 11973, USA}
\let\thefootnote\relax\footnote{$\ssymbol{3}$ Presently at: Korea Institute of Geoscience and Mineral Resources, Daejeon, 34132, Korea}
\let\thefootnote\relax\footnote{$\ssymbol{4}$ Presently at: NASA Marshall Space Flight Center, Huntsville, Alabama 35812, USA}
\addtocounter{footnote}{-1}\let\thefootnote\svthefootnote
}
\par\noindent

\end{sloppypar}

\section*{Acknowledgements}

The Telescope Array experiment is supported by the Japan Society for
the Promotion of Science(JSPS) through
Grants-in-Aid
for Priority Area
431,
for Specially Promoted Research
JP21000002,
for Scientific  Research (S)
JP19104006,
for Specially Promoted Research
JP15H05693,
for Scientific  Research (S)
JP19H05607,
for Scientific  Research (S)
JP15H05741,
for Science Research (A)
JP18H03705,
for Young Scientists (A)
JPH26707011,
for Transformative Research Areas (A)
JP25H01294,
for International Collaborative Research
24KK0064,
and for Fostering Joint International Research (B)
JP19KK0074,
by the joint research program of the Institute for Cosmic Ray Research (ICRR), The University of Tokyo;
by the Pioneering Program of RIKEN for the Evolution of Matter in the Universe (r-EMU);
by the U.S. National Science Foundation awards
PHY-1806797, PHY-2012934, PHY-2112904, PHY-2209583, PHY-2209584, and PHY-2310163, as well as AGS-1613260, AGS-1844306, and AGS-2112709;
by the National Research Foundation of Korea
(2017K1A4A3015188, 2020R1A2C1008230, and RS-2025-00556637) ;
by the Ministry of Science and Higher Education of the Russian Federation under the contract 075-15-2024-541, IISN project No. 4.4501.18, by the Belgian Science Policy under IUAP VII/37 (ULB), by National Science Centre in Poland grant 2020/37/B/ST9/01821, by the European Union and Czech Ministry of Education, Youth and Sports through the FORTE project No. CZ.02.01.01/00/22\_008/0004632, and by the Simons Foundation (MP-SCMPS-00001470, NG). This work was partially supported by the grants of the joint research program of the Institute for Space-Earth Environmental Research, Nagoya University and Inter-University Research Program of the Institute for Cosmic Ray Research of University of Tokyo. The foundations of Dr. Ezekiel R. and Edna Wattis Dumke, Willard L. Eccles, and George S. and Dolores Dor\'e Eccles all helped with generous donations. The State of Utah supported the project through its Economic Development Board, and the University of Utah through the Office of the Vice President for Research. The experimental site became available through the cooperation of the Utah School and Institutional Trust Lands Administration (SITLA), U.S. Bureau of Land Management (BLM), and the U.S. Air Force. We appreciate the assistance of the State of Utah and Fillmore offices of the BLM in crafting the Plan of Development for the site.  We thank Patrick A.~Shea who assisted the collaboration with much valuable advice and provided support for the collaboration’s efforts. The people and the officials of Millard County, Utah have been a source of steadfast and warm support for our work which we greatly appreciate. We are indebted to the Millard County Road Department for their efforts to maintain and clear the roads which get us to our sites. We gratefully acknowledge the contribution from the technical staffs of our home institutions. An allocation of computing resources from the Center for High Performance Computing at the University of Utah as well as the Academia Sinica Grid Computing Center (ASGC) is gratefully acknowledged.

}

\end{document}